\documentclass[reprint,aps,prl,showpacs,preprintnumbers,amsmath,amssymb,superscriptaddress]{revtex4-1}

\usepackage{graphicx}
\usepackage{dcolumn}
\usepackage{bm}
\usepackage{color}

\begin{document}


\title{Electron-phonon coupling and momentum-dependent electron dynamics in EuFe$_2$As$_2$ using time- and angle-resolved photoemission spectroscopy}

\author{L. Rettig}
\affiliation{Fachbereich Physik, Freie Universit\"at Berlin, Arnimallee 14, D-14195 Berlin, Germany}
\affiliation{Fakult\"at f\"ur Physik, Universit\"at Duisburg-Essen, Lotharstr. 1, D-47048 Duisburg, Germany}
\author{R. Cort\'es}
\affiliation{Fachbereich Physik, Freie Universit\"at Berlin, Arnimallee 14, D-14195 Berlin, Germany}
\affiliation{Abt. Physikalische Chemie, Fritz-Haber-Institut d. MPG, Faradayweg 4-6, D-14195 Berlin, Germany}
\author{S. Thirupathaiah}
\affiliation{Helmholtz-Zentrum Berlin, Albert-Einstein-Stra{\ss}e 15, D-12489 Berlin, Germany}
\author{P. Gegenwart}
\author{H.S. Jeevan}
\affiliation{I. Physik. Institut, Georg-August-Universit\"at G\"ottingen, D-37077 G\"ottingen, Germany}
\author{T. Wolf}
\affiliation{Karlsruhe Institute of Technology, Institut f\"ur Festk\"orperphysik, D-76021 Karlsruhe, Germany}
\author{U. Bovensiepen}
\email[Corresponding author: ]{uwe.bovensiepen@uni-due.de}
\affiliation{Fakult\"at f\"ur Physik, Universit\"at Duisburg-Essen, Lotharstr. 1, D-47048 Duisburg, Germany}
\affiliation{Fachbereich Physik, Freie Universit\"at Berlin, Arnimallee 14, D-14195 Berlin, Germany}
\author{M. Wolf}
\affiliation{Fachbereich Physik, Freie Universit\"at Berlin, Arnimallee 14, D-14195 Berlin, Germany}
\affiliation{Abt. Physikalische Chemie, Fritz-Haber-Institut d. MPG, Faradayweg 4-6, D-14195 Berlin, Germany}
\author{H.A. D\"urr}
\affiliation{Helmholtz-Zentrum Berlin, Albert-Einstein-Stra{\ss}e 15, D-12489 Berlin, Germany}
\affiliation{SLAC National Accelerator Laboratory, Menlo Park, CA 94025, USA}
\author{J. Fink}
\affiliation{Helmholtz-Zentrum Berlin, Albert-Einstein-Stra{\ss}e 15, D-12489 Berlin, Germany}
\affiliation{Leibniz-Institute for Solid State and Materials Research Dresden, P.O.Box 270116, D-01171 Dresden, Germany}

\date{\today}

\begin{abstract}
The Fe pnictide parent compound EuFe$_2$As$_2$ exhibits a strongly momentum dependent carrier dynamics around the hole pocket at the center of the Brillouin zone. The very different dynamics of electrons and holes cannot be explained solely by intraband scattering and interband contributions have to be considered. In addition, three coherently excited modes at frequencies of 5.6, 3.1 and 2.4 THz are observed.
An estimate of the electron-phonon coupling parameter reveals $\lambda<0.5$, suggesting a limited importance of e-ph coupling to superconductivity in Fe pnictides.
\end{abstract}


\maketitle

The new FeAs based high-$T_c$ superconductors (SC) exhibit a complex interplay between electronic, magnetic and lattice degrees of freedom~\cite{Mazin2010}. To understand superconductivity in these materials, one has to unravel the elementary excitations governing the semi-metallic ground state. Angle-resolved photoemission spectroscopy (ARPES) has analyzed the dressing of charge carriers in cuprate high-$T_c$ superconductors due to bosonic excitations by measuring the renormalization of the electronic states in the energy-momentum space. In FeAs SC, however, such an approach to study quasiparticle excitations has been less successful so far. Femtosecond (fs) time-resolved ARPES (trARPES) allows energy- and momentum-resolved investigation of the elementary relaxation processes directly in the time domain.

\begin{figure}[tb]
\includegraphics[width=8.6cm]{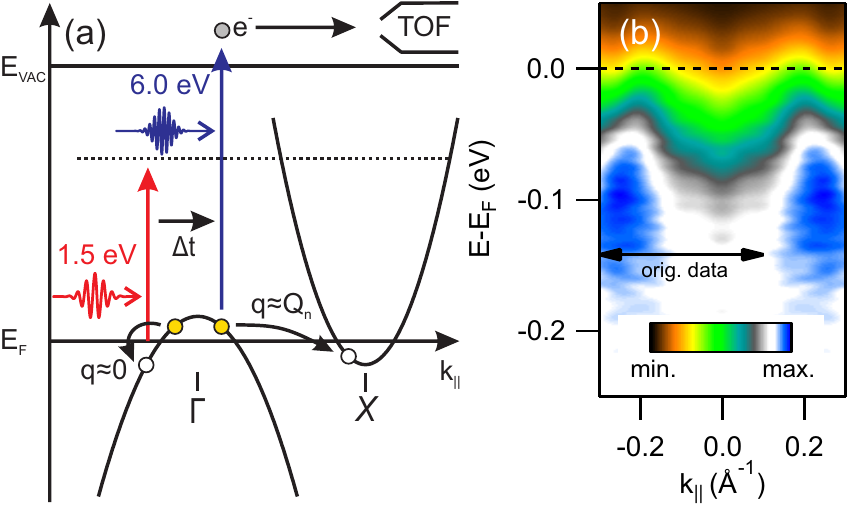}  
\caption{
\label{fig:fig1}
(color online) (a) Schematic electron band structure of Fe pnictides along $\Gamma$-$\textit{X}$ and sketch of the pump-probe experiment. Electrons are excited into weakly dispersing states and create e-h pairs in both bands. Black arrows indicate relaxation pathways (see text). The probe photon energy of 6.0 eV allows for detection of states near $\Gamma$. Photoelectrons are analyzed by a time-of-flight spectrometer (TOF). (b) Laser ARPES spectra of EuFe$_2$As$_2$ taken at $T$=100 K without optical excitation. Data have been symmetrized around $\Gamma$.
}
\end{figure}

The electronic bands forming the Fermi surface and determining the low-energy excitations consist in Fe pnictides of hole pockets at the center of the Brillouin zone (BZ) ($\Gamma$-point) and electron pockets at the zone corner ($\textit{X}$-point, see Fig.~\ref{fig:fig1}(a))~\cite{Singh2008,Liu2008}.
Spin fluctuations related to Fermi surface nesting are considered as one origin for Cooper pair formation~\cite{Mazin2008a}. 
As the hole (electron) pockets are almost filled (empty) in these systems, small structural deformations or doping will shift the Fermi level into a region of heavier carriers~\cite{Yin2008a} and may change the nesting conditions. 
This provides a scenario for the sensitivity of $T_c$ to the lattice structure.
It was discussed that the Fe-As-Fe bonding angle~\cite{Lee2008} or the Fe-As distance~\cite{Kuroki2009} is important for $T_c$ and the symmetry of the superconducting order parameter.

Another possible pairing mechanism is electron-phonon (e-ph) interaction. On the one hand, calculations for LaFeAsO$_{1-x}$F$_x$ and BaFe$_2$As$_2$ resulted in weak e-ph coupling~\cite{Boeri2008, Boeri2010}. On the other hand, stronger coupling was predicted for particular modes which distort the Fe-As tetrahedra and thus change the nesting conditions. These are the Raman active $A_{1g}$ mode at the zone center corresponding to a breathing mode of As ions perpendicular to the Fe layers with an energy of 22~meV and an opposite vibration of the Fe and As ions parallel to the FeAs layers with a wave vector satisfying the nesting condition $q=Q_n=\Gamma-X\equiv(\pi,\pi)$.

In this letter we study the fs dynamics of the parent compound EuFe$_2$As$_2$ ($T_N\approx190\,\textrm{K}$)~\cite{Jeevan2008} using trARPES to analyze the interaction of charge carriers with low-energy excitations of the material~\cite{Rhie2003,Perfetti2007,Schmitt2008}. We observe a strong momentum anisotropy in the electron-electron (e-e) and electron-phonon (e-ph) relaxation, driven by intra- and interband scattering, as well as coherent modes. 

We employ ultrashort (55~fs), intense infrared (h$\nu_1$=1.5~eV) laser pulses for optical excitation. A time-delayed ultraviolet (h$\nu_2$=6.0~eV, 80~fs) pulse monitors the energy and momentum dependent single-particle spectral function as a function of pump-probe delay by ARPES as sketched in Fig.~\ref{fig:fig1}(a). The energy resolution of $50\,\textrm{meV}$ is determined by the spectrometer and the bandwidth of the probe pulses. The  momentum resolution is $0.05\,\textrm{\AA}^{-1}$ and the temporal resolution is $<$100~fs, for details see~\cite{Schmitt2008}. Single crystals of EuFe$_2$As$_2$ grown in Sn-flux~\cite{Jeevan2008} were cleaved in ultrahigh vacuum at $T$=100~K, where all measurements were carried out, if not otherwise stated. Spectra of different emission angles have been normalized at $E-E_F=-0.4\,\textrm{eV}$ to account for polarization and matrix element effects. 

Laser ARPES intensity of EuFe$_2$As$_2$ obtained at h$\nu$=6.0~eV without optical excitation is shown in Fig.~\ref{fig:fig1}(b) as a function of energy and momentum. The observed hole pocket around the $\Gamma$-point is comparable to data taken at higher photon energies~\cite{deJong2010}. 

\begin{figure}[tb]
\includegraphics[width=8.6cm]{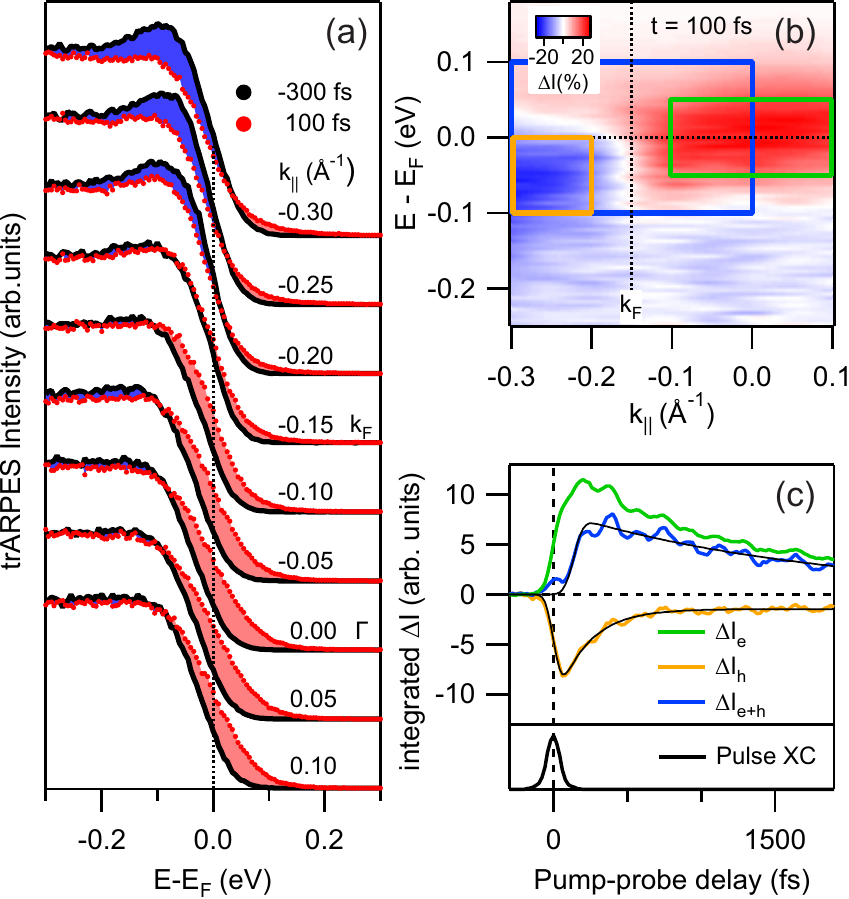}  
\caption{
\label{fig:fig2}
(color online) (a) Time-resolved trARPES spectra of EuFe$_2$As$_2$ for different parallel electron momenta before and after excitation with an absorbed fluence of $F=280\,\mu\textrm{J}/{\textrm{cm}}^{2}$. Depletion and increase of spectral weight $\Delta I$ is marked by blue and red areas, respectively. Spectra are vertically offset for clarity. (b) Color map of $\Delta I$ for $t$=100~fs. Boxes mark integration areas for $\Delta I$ shown in (c). (c) Upper panel: delay dependent $\Delta I$, integrated over energy and momentum intervals according to (b), sampling hole dynamics for $E<E_F$ (yellow curve) and electron dynamics for $E>E_F$ (green curve) in the vicinity of the hole pocket. The total low-energy intensity $\Delta I_{e+h}$ (blue curve) shows a delayed increase at $140\,\textrm{fs}$ after excitation. All curves have been normalized to the integration area. Black lines are exponential fits to the data. Lower panel: cross-correlation (XC) trace obtained from electrons with $E-E_F > 1\,\textrm{eV}$.}
\end{figure}

Spectra for different pump-probe delays are shown in Fig.~\ref{fig:fig2}(a) as energy distribution curves (EDCs) before (black) and 100~fs after excitation (red). The pump induced changes are emphasized by blue (depletion) and red (increase) areas. This difference $\Delta I(E,k,t) = I(E,k,t) - I_0(E,k)$ is shown in a color coded intensity plot as function of energy and momentum in Fig.~\ref{fig:fig2}(b). Two distinct types of response are found: near the $\Gamma$-point within the hole pocket, an increase in spectral weight is observed (red), whereas we find a decrease for occupied states (blue). This behavior is the fingerprint of electron-hole (e-h) excitations, leading to a transfer of electrons from occupied states outside the hole pocket into empty states inside the pocket.

Integrating the data within representative energy/momentum intervals (Fig.~\ref{fig:fig2}(b)) yields the temporal evolution of electron and hole population, $\Delta I_e$ and $\Delta I_h$, depicted in Fig.~\ref{fig:fig2}(c). We find (i) a different sign of the response and (ii) different timescales of buildup and decay for the electron and the hole population. The number of holes initially follows the pump-pulse integral and peaks at $60\,\textrm{fs}$. The subsequent decay is fitted with a single-exponential decay convoluted with the laser pulse cross-correlation, yielding $\tau_{h}=220(10)\,\textrm{fs}$. After $750\,\textrm{fs}$, the excitation has relaxed to a value slightly higher than initially, indicating an elevated lattice temperature. In contrast, the electron population shows a stronger increase, which continues to rise after $100\,\textrm{fs}$, when the pump pulse is over. Integration over electrons and holes near $\Gamma$ (blue box in Fig.~\ref{fig:fig2}(b)) yields $\Delta I_{e+h}$, which compares well to the sum $\Delta I_e + \Delta I_h$ and reveals a transient increase of the total number of electrons near $\Gamma$ with a delayed onset of $140\,\textrm{fs}$ with respect to time zero. This excess electron population decays with a time constant $\tau_{e}=1760(60)\,\textrm{fs}$, clearly slower than $\tau_h$. 

We now discuss the origin of this asymmetry of electron and hole dynamics. Electrons excited by the laser pulse to an unoccupied band (Fig.~\ref{fig:fig1}(a)) quickly relax by e-e scattering, thereby exciting additional electrons. Such intraband excitation processes create e-h pairs in the same band and thereby lead to an equal number of excited electrons and holes within the one band~\cite{Lisowski2004}. This is observed during the first $\sim100\,\textrm{fs}$. Later, more excited electrons than holes are detected which originate from different bands of the BZ, e.g. the bands forming the electron pocket at the $\textit{X}$-point (Fig.~\ref{fig:fig1}(a)). The required momentum vector $Q_n$ for this interband scattering from $\textit{X}$ to $\Gamma$ can be provided by defect, spin or phonon excitations, which might lead to additional fingerprints in the transient spectral function, discussed below. We explain this carrier imbalance by the enhanced density of states for electrons and holes at $\Gamma$ and $\textit{X}$, respectively~\cite{Singh2008}, leading to an effective electron transfer from $\textit{X}$ to $\Gamma$, which could be enhanced by nesting.

Now we turn to the subsequent relaxation of electrons and holes. It is mediated by e-ph scattering or spin excitations, leading to e-h recombination and energy transfer to the lattice or spin subsystem. Due to the e-h imbalance, part of the e-h pairs will recombine by scattering with $q\approx0$ modes within the $\Gamma$ band (Fig.~\ref{fig:fig1}(a), left). The remaining excess population of electrons has to recombine with holes at $\textit{X}$ generating excitations with large $q$ near $Q_{n}$ (Fig.~\ref{fig:fig1}(a), right). From the longer lifetime of the excess electron population at $\Gamma$ we thus conclude a lower scattering probability for the latter process. 

\begin{figure}[tb]
\includegraphics[width=8.6cm]{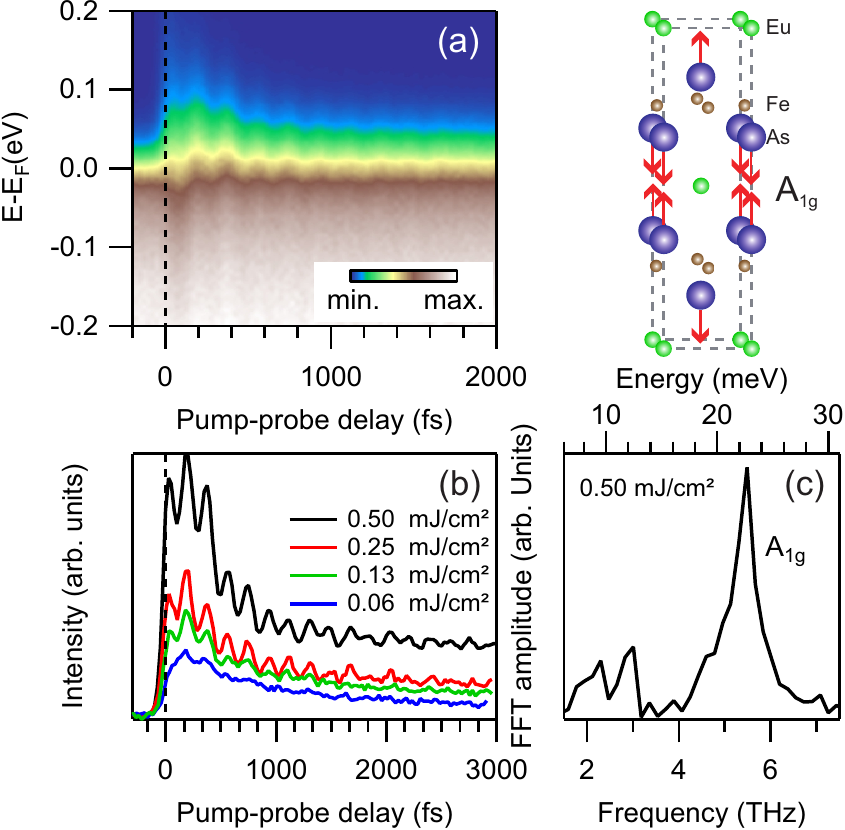}  
\caption{
\label{fig:fig3}
(color online) (a) trARPES intensity as a function of energy and pump-probe delay taken at $\Gamma$ with $F$=0.5~mJ/cm$^{2}$. (b) trARPES intensity integrated over an energy window of $1.5\,\textrm{eV}$ above $E_F$ for various $F$. (c) Fast Fourier transform of the $F$=0.5~mJ/cm$^{2}$  data in (b). The sketch shows the nuclear motion corresponding to the $A_{1g}$ mode at $5.6\,\textrm{THz}$.
}
\end{figure}

A closer look at Fig.~\ref{fig:fig2}(c) reveals a periodic modulation superimposed on the decay. To investigate these oscillations, stronger excitation has been used. Fig.~\ref{fig:fig3}(a) shows the trARPES intensity at the $\Gamma$-point for $F$=0.5~mJ/cm$^{2}$. As function of delay, pronounced oscillations modulate the photoemission signal around $E_F$. Analyzing the spectral cutoff near $E_F$ (Fig.~\ref{fig:fig2}(a)) shows that the distribution function itself is modulated, leading to a transient oscillation of its position~\cite{SOM}. Hence, these low-energy excitations couple efficiently to states at $E_F$.

The time dependent intensity integrated between $E_F$ and $E-E_F=1.5\,\textrm{eV}$ is depicted in Fig.~\ref{fig:fig3}(b) for various fluences. Within the first $2\,\textrm{ps}$ after excitation, a strong oscillation mode with a frequency of $5.6\,\textrm{THz}$ is observed. For $t>1.5\,\textrm{ps}$, an interference pattern is found which indicates further frequencies. Next, the incoherent contribution is subtracted~\cite{Melnikov2008}, and frequencies are determined by fast Fourier transformation (FFT). Fig.~\ref{fig:fig3}(c) depicts the normalized FFT  presenting a sharp peak at h$\nu_1$=23(1)~meV and two weaker modes at h$\nu_2$=13(1)~meV and h$\nu_3$=10(1)~meV. All three modes have been found with similar frequencies in doped and undoped samples of the BaFe$_2$As$_2$ family~\cite{SOM}, suggesting that these coherent excitations are a more general phenomenon in 122 Fe pnictides. The frequencies of all modes are found to be independent of fluence or temperature up to $T=300~\textrm{K}$.

By comparison with Raman scattering~\cite{Litvinchuk2008} we identify the mode at 23 meV with the $A_{1g}$ phonon mode, corresponding to a displacement of the As atoms perpendicular to the Fe layers (Fig.~\ref{fig:fig3}). Coherent excitation of this mode has also been observed in time-resolved reflectivity~\cite{Mansart2009}. The excitation of this fully symmetric mode in a semi-metal can be described by the displacive excitation of coherent phonons (DECP) model~\cite{Zeiger1992}. The observation of coherent phonons in trARPES requires strong coupling of the vibrations to electronic states. Hence our observation is consistent with calculations of LaFeAsO$_{1-x}$F$_x$ ~\cite{Boeri2008}, which predicts enhanced e-ph coupling for the $A_{1g}$ mode. 

The assignment of the other two modes is less clear. The DECP mechanism allows to first order only excitation of fully symmetric modes of $A_1$ symmetry~\cite{Zeiger1992}. However, the coherent excitation of non fully symmetric modes has been detected in the bulk~\cite{Hase2005} and at surfaces~\cite{Melnikov2008} of metals. Calculations~\cite{Boeri2008} predict modes at the zone-center around 10 meV, which are, however, not Raman active and couple only weakly to the electronic system. The calculations show a maximum in the Eliashberg spectral function $\alpha^2F(\omega)$ around 13~meV due to modes at the BZ boundary, probably related to Fe-As shear mode vibrations with $q=Q_n$. Coherent excitation of phonon modes with finite momenta is usually prohibited, but the ultrafast redistribution of electrons in the BZ (Fig.~\ref{fig:fig2}) could provide the required momentum. Furthermore, contributions of surface phonons or spin excitations~\cite{Melnikov2008, Inosov2010} cannot be ruled out.

Our observation demonstrates coupling of coherent modes, as we detect their imprint on electronic states near $E_F$. To quantify the e-ph coupling, we use the energy resolution provided by trARPES. Usually, e-ph relaxation is described using the two temperature model (2TM)~\cite{Allen1987,Perfetti2007}. However, the different dynamics of electrons and holes contradict the assumption of a thermalized electronic subsystem, which questions the application of the 2TM. Here, we overcome this limitation by analyzing the energy relaxation of excited electrons directly.

\begin{figure}[tb]
\includegraphics[width=8.6cm]{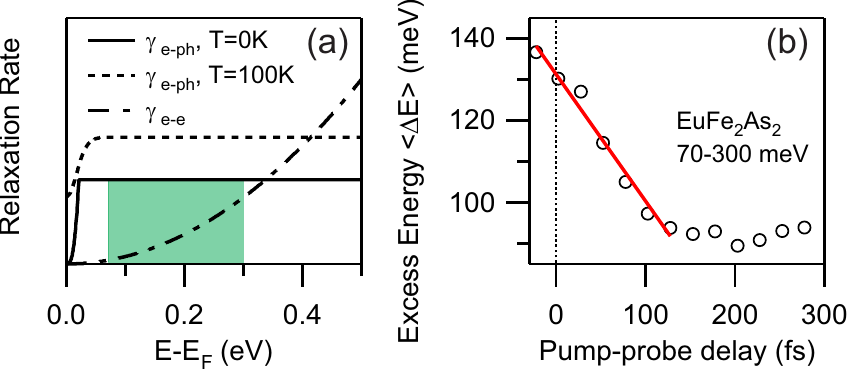}  
\caption{
\label{fig:fig4}
(color online) (a) Electron-phonon (e-ph) and electron-electron (e-e) contributions to the decay rate. The e-ph part is given for $T=0\,\textrm{K}$ and $T=100\,\textrm{K}$. The shaded area marks the energy where e-ph scattering dominates. (b) Electronic mean excess energy extracted from trARPES data near $k_F$ within the energy window marked in (a). The solid line is a fit to eq.~\eqref{eqn:e-ph}.}
\end{figure}

The scattering rate of electrons excited at energy $\epsilon=E-E_F$ is determined by $\gamma = \hbar\tau^{-1} = 2 \Im\Sigma(\epsilon)$, where $\Sigma$ is the electronic self energy. Important energy dependent contributions to $\gamma$ arise from e-e and e-ph scattering. e-e scattering is considered to follow Fermi liquid theory, $\gamma_{e-e}\propto\epsilon^2$ (Fig.~\ref{fig:fig4}(a), dash-dotted line). At low excitation energies, this contribution is small compared to the e-ph scattering, which increases up to the maximal phonon energy $\hbar\omega_{max}$ and is constant above (Fig.~\ref{fig:fig4}(a)). For $\epsilon>\hbar\omega_{max}$ and $T=0\,\textrm{K}$, $\gamma_{e-ph}$ results for an Einstein mode $\omega$ to 
$\gamma_{e-ph}=\pi\hbar\lambda\omega$~\cite{Engelsberg1963}, where $\lambda$ is the e-ph coupling constant. Within the energy window between $\hbar\omega_{max}$ and the crossover regime, where $\gamma_{e-e}$ becomes dominant (shaded area in Fig.~\ref{fig:fig4}(a)), the rate of energy dissipation due to the emission of a phonon with energy $\hbar\omega$ is given by 
\begin{equation}
\label{eqn:e-ph}
\frac{\textrm{d} E}{\textrm{d} t} = \frac{\hbar\omega}{\tau}=\pi\hbar\lambda\omega^2\qquad,
\end{equation}
which leads to a linear decay. This rate of energy relaxation can be extracted from the experimental data by analyzing the mean excess energy, $\left\langle\Delta E(t)\right\rangle = \int^{\epsilon_1}_{\epsilon_0}{\epsilon \Delta I(\epsilon,t)\textrm{d}\epsilon}/\int^{\epsilon_1}_{\epsilon_0}{\Delta I(\epsilon,t)\textrm{d}\epsilon}$. Data of EuFe$_2$As$_2$ near $k_F$ are shown in Fig.~\ref{fig:fig4}(b). To minimize lattice heating and e-e scattering, a low $F\sim50\,\mu\textrm{J}/\textrm{cm}^2$ has been used and the integration window of $\epsilon=70-300$~meV is carefully chosen according to the conditions discussed above. The linear fit yields according to \eqref{eqn:e-ph} a value of $\lambda\omega^2=70\,\textrm{meV}^2$. Since e-e scattering cannot be totally neglected and finite temperatures lead to faster relaxation, see Fig.~\ref{fig:fig4}(a), this value provides an upper limit. Our results for $\lambda\omega^2$ agree well with recently published results using optical pump-probe schemes \cite{Mansart2010, Stojchevska2010}, which however rely on the 2TM or similar models to extract $\lambda$. Comparing the timescales of the relaxation in Fig.~\ref{fig:fig4} and Fig.~\ref{fig:fig2} shows, that mostly fast intraband transitions with $q\approx0$ dominate the energy relaxation.

Based on this we estimate the value of $\lambda$ for a particular mode. Considering the coherently excited $A_{1g}$ mode at 23 meV, which is strongly coupled, we find $\lambda<0.2$. This estimate is in agreement with calculations \cite{Boeri2008, Boeri2010} of various Fe-pnictide compounds.
Even if we consider the lowest coupled modes to be most important for e-ph coupling, $\lambda$ does not exceed 0.5. Similar small values for $\lambda$ have been found in the cuprate high-$T_c$ SC \cite{Perfetti2007}, suggesting limited importance of e-ph coupling for the pairing mechanism in both classes of materials.

In conclusion, using trARPES we investigated intraband and interband e-e scattering, coherent low-energy excitations and e-ph energy relaxation in the Fe pnictide parent compound EuFe$_2A$s$_2$. We find different electron and hole dynamics, which is explained by interband transitions between $\Gamma$ and $\textit{X}$ leading to redistribution of carriers between the respective bands. In Eu and Ba 122 pnictides, three specific coherent modes couple directly to electronic states close to the Fermi level. The analysis of e-ph coupling reveals $\lambda<0.5$ which makes e-ph coupling an unlikely pairing candidate for superconductivity.

\begin{acknowledgments}
We acknowledge support in data evaluation by A. Melnikov. This work has been funded by the DFG through BO 1823/2 and SPP 1458. R.C. acknowledges the Alexander von Humboldt Foundation.
\end{acknowledgments}

%

\end{document}